# Explicit Construction of First Integrals by Singularity Analysis in Nonlinear Dynamical Systems

**Christos Efthymiopoulos**[*], **Tassos Bountis**[†] **and Athanasios Manos** [†]

[*] Center for Astronomy and Applied Mathematics,
Academy of Athens

[†] Center for Research on Applied Nonlinear Systems (CRANS)
Department of Mathematics, University of Patras

**Keywords:** *Singularity analysis, Painlevé test, Computation of First Integrals*

**Abstract.** *The Painlevé and weak Painlevé conjectures have been used widely to identify new integrable nonlinear dynamical systems. The calculation of the integrals relies though on methods quite independent from Painlevé analysis. This paper proposes a new explicit algorithm to build the first integrals of a given set of nonlinear ordinary differential equations by exploiting the information provided by the Painlevé - Laurent series representing the solution in the neighbourhood of a movable singularity. The algorithm is based on known theorems from the theory of singularity analysis. Examples are given of the explicit construction of the first integrals in nonlinear Hamiltonian dynamical systems with a polynomial potential, and in generalized Volterra systems.*

## 1 INTRODUCTION

A systematic search for first integrals in dynamical systems starts usually with the implementation of the so-called 'direct' method. This method assumes a particular functional form of the integral, e.g., polynomial, rational or algebraic. Terms of the same form are combined in a function, which has, presumably, a zero time derivative. This turns the problem to finding a proper balancing of the coefficients of the various terms so that the selected function represents an integral.

The direct method proved fruitful in discovering the integrals in various types of nonlinear dynamical systems. Of particular interest is the case of autonomous systems described by equations of the form

$$\dot{x}_i = F_i(x_1, x_2, ..., x_n), \quad i=1,2,...,n \qquad (1)$$

where the functions $F_i$ are polynomial, rational or algebraic. For these equations, integrals of the form

$$\Phi = \sum_{k_i} a_{k_1, k_2, ..., k_n} \prod_{i=1}^{n} x_i^{k_i} \qquad (2)$$

where the exponents $k_i$ are natural numbers (polynomial first integrals), integers (rational first integrals), or rational (algebraic first integrals) can be recovered by the direct method. A particular example refers to Hamiltonian systems with a polynomial potential. There, applications of the direct method lead to new discoveries of integrable systems, and even to the production of tables of integrable systems (Hietarinta 1983, 1987, Dorizzi et al. 1986, Grammatikos et al.1983, Cleary 1989, Nakagawa and Yoshida 2001).

Failure to find an integral of the form (2), of some maximum degree, does not imply non-existence of an integral in general for the given set of ODEs. In principle, the search should be extended to integrals of an arbitrarily high maximum degree. This is a hardly tractable task even with the use of computer algebraic manipulators. Fortunately, there are useful criteria to restrict in advance the maximum degree of search for an integral of the form (2). Such criteria were built on the ground of *singularity analysis* for systems of the form (1); hence they are applicable to systems possessing at least the *weak Painlevé* property (Ramani et al. 1982). Since the latter is a necessary condition for algebraic integrability (Yoshida 1983a), the relevant criteria apply to a wide class of systems possessing integrals of the form (2). Precisely, there is a number of theorems, which establish a connection between the so-called *resonances,* found by singularity analysis of the set of ODEs, and the degree of a first integral associated with the ODEs. Such are the theorems of Yoshida (1983b), Furta (1996), Goriely (1996), Tsygvintsev (2001) and Llibre and Zhang (2002).

The purpose of the present paper is to present a new algorithm able to recover first integrals of the form (2) in systems of the form (1), which is based purely on a singularity analysis of the corresponding ODEs. The theorem of Yoshida (1983b) is used to calculate upper limits for the exponents $k_i$ appearing in Eq.(2). Then, the information contained in *the Painlevé - Laurent series,* constructed by singularity analysis, is exploited in order to recover the integrals algorithmically. Technically, this is a problem of balancing of coefficients similar to the problem of balancing of coefficients in the direct method. The problem is solved by means of the singular value decomposition technique. Preliminary notions needed in the implementation of the algorithm are given in section 2. The algorithm is exposed in section 3, where its application is illustrated by specific examples. Section 4 is a summary of the main conclusions. It should be stressed that the Painlevé - Laurent series are the *only* input in the algorithm. This fact implies that the information on the integrals is contained in the solutions given as Painlevé - Laurent series in the neighborhood of a movable singularity of the ODEs (1). The implication of the last remark to the problem of the connection between integrability and singularity analysis is an open issue of current research.

## 2   CONSTRAINTS ON THE FIRST INTEGRALS BY SINGULARITY ANALYSIS

Following Yoshida (1983b), a system of ODEs of the form (1) is called scale-invariant if there are exponents $\lambda_i$ such that the equations (1) remain invariant under the scale transformation $x_i \to a^{\lambda_i} x_i$, $t \to a^{-1} t$. It follows that a scale-invariant system (1) admits exact particular solutions of the form

$$x_i = \frac{d_i}{\tau^{\lambda_i}}, \tag{3}$$

where the set of $d_i$ is any root of the system of algebraic equations $F_i(d_1, d_2, ..., d_n) - \lambda_i d_i = 0$, and $\tau = t - t_0$ where $t_0$ is a movable (initial condition-dependent) singularity in the complex time plane. One set of $d_i$, and the associated exponents $\lambda_i$, are called a *balance*. In the sequel, the exponents $\lambda_i$ are assumed to be rational numbers.

The singularity analysis of the ODEs (1) can be implemented by the Ablowitz - Ramani - Segur (ARS) algorithm (Ablowitz et al. 1980). In this algorithm, the special solution (3) arises as one of the so-called 'dominant terms', for which $d_i \neq 0$ for all $i=1,...,n$. However, there may also be other types of dominant terms, with different exponents $\lambda_i$. These correspond to special solutions of the form (3) where $d_i = 0$ for one or more $i$. In both cases, the dominant terms are the leading terms of a series solution, which can be developed in the neighborhood of the singularity $t_0$.

The second step in the ARS algorithm is to construct the series, called `Painlevé - Laurent series', for all the balances. These are Laurent series when $\lambda_i$ are integers, otherwise they are series in rational powers of $\tau$ (called Puiseux series). To build up one of the series, first, its *resonances* are specified. These are the values of $r$ for which the coefficients of the terms $\tau^{r-\lambda_i}$ in the series are arbitrary. In the case of a series corresponding to a balance with $d_i \neq 0$ for all $i=1,...,n$, the resonances are equal to the eigenvalues of the matrix $K_{ij} = \partial F_i / \partial x_j - \delta_{ij} \lambda_i$ and they are called Kowalevski exponents (Yoshida, 1983a). For other types of balances, the corresponding resonances are in general not equal to the Kowalevski exponents defined as above, but they differ from them by a constant exponent (Ramani et al. 1988).

Now, following Yoshida, a function $\Phi$ of the variables $x_i$ will be called *weighted-homogeneous* of weight $M$, with respect to the exponents $\lambda_i$, if the following property holds:

$$\Phi(a^{\lambda_1} x_1, a^{\lambda_2} x_2, ..., a^{\lambda_n} x_n) = a^M \Phi(x_1, x_2, ..., x_n) \tag{4}$$

Then we have the following

**Theorem 1** Yoshida's (1983b): *if a scale-invariant system of ODEs of the form (1) has a weighted - homogeneous first integral $\Phi$, of weight degree $M$, and the elements of $\nabla \Phi(d_i)$ for the balance $d_i$ are finite and non-zero, then $M$ becomes one of the Kowalevski exponents associated with this balance.*

It follows immediately that the Painlevé - Laurent series for the balance $d_i$ can be written as:

$$x_i = x_{iE} + R_i = \frac{d_i}{\tau^{\lambda_i}} + ... + A_i \tau^{rmx - \lambda_i} + R_i \tag{5}$$

where *rmx*, the maximum resonance, satisfies the relation $rmx \geq M$. The part denoted as $x_{iE}$ will be called the essential part of the Painlevé - Laurent series (5), while $R_i$, called the remainder, contains the infinite number of terms in the series of powers $\tau^q$ with $q > rmx - \lambda_i$.

Consider now a weighted-homogeneous first integral $\Phi$ of the form (2). Then, the exponents $k_i$ satisfy the relation

$$\sum_{i=1}^{n} k_i \lambda_i = M \qquad (6)$$

due to (4). In the most general case, $\lambda_i = p/\rho$, where $p, \rho$ are prime integers. If $\rho = 1$ the system is of the Painlevé type, while, if $\rho > 1$, the system is of the weak-Painlevé type. The remainder $R_i$ starts with terms of degree $O(\tau^{rmx-\lambda_i+1/p})$. It follows that the contribution of $R_i$ in $x_i^{k_i}$ is in terms of degree $O(\tau^{-\lambda_i k_i + rmx + 1/p})$ or higher. This means that the contribution of all the remainders in the expression (2) is in terms of degree $O(\tau^{-\sum \lambda_i k_i + rmx + 1/p}) = O(\tau^{-M + rmx + 1/p})$ or higher. But $rmx \geq M$. Thus, the remainders $R_i$ contribute only to terms of *positive* degree of $\tau$ in the expression (2).

On the other hand, by substituting the infinite series (5) into (2), all the terms of non-zero power of $\tau$ must vanish if the expression (2) represents an integral. Precisely, this is the key property to build the integrals from singularity analysis.

Since the $R_i$'s contribute only to terms of positive order, the substitution of only the essential parts $x_{iE}$ into (2) suffices to eliminate all terms $O(1/\tau^q)$ with $q > 0$. Hence we have the following

**Proposition 1:** *If a scale - invariant system (1) admits Painlevé - Laurent series solutions, and $\Phi$ is a weighted - homogeneous integral of (1), then the functional expression $\Phi(x_{1E}, x_{2E}, ..., x_{nE})$, where $x_{iE}$ is the essential parts of a Painlevé Laurent series, does not contain singular terms, i.e. terms $O(1/\tau^q)$ with $q > 0$.*

Let us note that the expression $\Phi(x_{1E}, x_{2E}, ..., x_{nE})$ is not a constant, i.e. time independent. Proposition 1 asserts only that this expression is not singular as the time t approaches a movable singularity in a complex time domain.

Dropping the assumption of scale invariance of the system (1), we consider systems (1) where the functions $F_i$ are given by sums of terms:

$$F_i = F_i^{(1)} + F_i^{(2)} + ... + F_i^{(m_i)} \qquad (7)$$

where the functions $F_i^{(j)}$ are homogeneous of degree j, $m_i$ are rational and $m_i \geq j$ for all $j$. Then, if the system (1) has Painlevé - Laurent type solutions, one of its dominant behaviors and associated resonances is determined by the homogeneous term $F_i^{(m_i)}$ alone, i.e. this solution will have the same dominant terms (i.e. same $\lambda_i$) and the same resonances as the system:

$$\dot{x}_i = F_i^{(m_i)}(x_1, x_2, ..., x_n), \quad i = 1, 2, ..., n. \qquad (8)$$

The full system has Painlevé - Laurent type solutions provided that the functions $F_i^{(j)}$, $j < m_i$ satisfy certain compatibility conditions (Ablowitz et al. 1980).

If $I$ is an integral of the system expressed as sum of weighted - homogeneous terms, then its terms of maximum degree will be defined by the scale - invariant system (8), i.e. they will be of the form (2), with $k_i$ satisfying (6) (Hietarinta 1983, Nakagawa and Yoshida 2001, Nakagawa 2002). Any other term of this integral will be of the form $a_{k_1', k_2', ..., k_n'} \prod_{i=1}^{n} x_i^{k_i'}$ with $\sum k_i' < \sum k_i$. One can easily verify that if a term $a_{k_1', k_2', ..., k_n'} \prod_{i=1}^{n} x_{Ei}^{k_i'}$ does not introduce terms singular in $\tau$, then neither does so the term $a_{k_1', k_2', ..., k_n'} \prod_{i=1}^{n} x_{Ei}^{k_i'}$.

Thus we arrive to the following:

**Proposition 2:** *If the functions $F_i$ and an integral $I$ of (1) are sums of homogeneous and weighted - homogeneous functions respectively, and (1) admits Painlevé - Laurent series solutions, then the expression $I(x_{1E}, x_{2E}, ..., x_{nE})$, where $x_{iE}$ is the essential part of the solution (5) associated to (8), does not contain terms $O(1/\tau^q)$ with $q > 0$.*

## 3 EXPLICIT CONSTRUCTION OF FIRST INTEGRALS

### 3.1 The Algorithm

Proposition 2 can be used to provide constraints in the exponents $k_i$ of the terms of an integral of the form (2) in systems with the Painlevé or weak Painlevé property. Thus the following algorithm can be used to determine a first integral explicitly by using the information from the Painlevé - Laurent series solutions.

**Step 1:** Suppose the existence of a first integral of the form (2) with undetermined coefficients $a_{k1,k2,...,kn}$.

**Step 2:** Do singularity analysis of the equations (1) and find all possible types of Painlevé - Laurent series solutions of the form (5).

**Step 3 (selection rule}:** Use Proposition 2 to determine the maximum possible value of the exponents $k_i$ in any term of (2). Thus, write $\Phi$ as a sum over a finite number of terms of the form (2).

**Step 4:** Substitute one of the Painlevé - Laurent expressions (5) into (2). Consider as free parameters the coefficients $a_{k1,k2,...,kn}$. Finally, find the values of the coefficients $a_{k1,k2,...,kn}$ for which $\tau$ *is eliminated from the expression (2)*.

This leads to an infinite homogeneous system of linear equations for the unknown coefficients $a_{kl}$, since there are infinite terms corresponding to an arbitrarily high power of $\tau$ that must be eliminated in the expression (2). In practice, however, we determine a finite homogeneous system of the form:

$$C \cdot A = 0 \tag{9}$$

where the matrix $C$, with constant coefficients, has dimension $M \times N$, with $M>N$, and $N$ is equal to the number of unknown coefficients $a_{k1,k2,...,kn}$. The column vector $A$, of length $N$, has the unspecified coefficients $a_{k1,k2,...,kn}$ as components. Setting the l.h.s. of Eq(9) equal to zero expresses the fact that the time $\tau$ is eliminated from the expression (2) up to terms of order $O(\tau^M)$. Thus, at the limit $M \to \infty$, any vector $A$ satisfying (9) represents a first integral of the form (2) with coefficients given by $A$. A complete set of basis vectors of the null space of $C$ represents a complete set of integrals of the form (2). For any finite $M$, this set is found by singular value decomposition. Then, it is checked a posteriori, by direct differentiation, that the expressions found represent indeed exact integrals.

## 3.2 First Example: Construction of the integrals in a 2D autonomous Hamiltonian system

We consider the Hamiltonian

$$H = \frac{1}{2}(p_x^2 + p_y^2 + x^2 + y^2) - x^2 y - 2y^3 \tag{10}$$

This system is both Painlevé and integrable (Bountis et al. 1982). The general solution for $x, y, p_x, p_y$ can be expressed locally, around a movable singularity, as a Laurent series with four arbitrary constants, namely:

$$\begin{aligned}
x(\tau) &= \frac{A}{\tau} + a_1\tau + B\tau^2 + a_3\tau^3 + a_4\tau^4 a_5\tau^5 + O(\tau^6) \\
y(\tau) &= \frac{1}{\tau^2} + b_0 + b_0\tau^2 + b_3\tau^3 + C\tau^4 + O(\tau^5) \\
p_x(\tau) &= -\frac{A}{\tau^2} + a_1 + 2B\tau + 3a_3\tau^2 + 4a_4\tau^3 5a_5\tau^4 + O(\tau^5) \\
p_y(\tau) &= -\frac{2}{\tau^3} + 2b_2\tau + 3b_3\tau^2 + 4C\tau^3 + O(\tau^3)
\end{aligned} \tag{11}$$

where $A, B, C$ together with $t_0 = t - \tau$ are arbitrary, and $b_0 = (1-A^2)/12$, $a_1 = A(1-2b_0)/2$, $b_2 = (b_0 - 6b_0^2 - 2A_1)/10$, $a_3 = (2b_0 a_1 - a_1 + 2Ab_2)/4$, $b_3 = -AB/3$, $a_4 = (2Ab_3 - B - 2Bb_0)/10$, $a_5 = (2AC + 2a_1 b_2 + 2b_0 a_3 - a_3)/18$.

We look for a second integral independent of and in involution with the Hamiltonian (10) in the polynomial form:

$$\Phi(x, p_x, y, p_y) = \sum_{k,l,m,n} a_{klmn} x^k p_x^l y^m p_y^n \tag{12}$$

Applying the selection rule given by Proposition 2 yields exactly 38 monomials in which substitution of (11) into (12) implies that the remainders of τhe four Laurent series contribute only to terms of positive order in $\tau$. The free coefficients in (12) are:

$$a_{0001}, a_{0010}, a_{0100}, a_{1000}, a_{0002}, a_{0011},$$
$$a_{0020}, a_{0101}, a_{0110}, a_{0200}, a_{1001}, a_{1010},$$

$$a_{1100}, a_{2000}, a_{0030}, a_{0120}, a_{0210}, a_{0300},$$
$$a_{1001}, a_{1020}, a_{1101}, a_{1110}, a_{1200}, a_{2001},$$
$$a_{2010}, a_{2100}, a_{3000}, a_{2020}, a_{2110}, a_{2200},$$
$$a_{3001}, a_{3010}, a_{3100}, a_{4000}, a_{4010}, a_{4100},$$
$$a_{5000}, a_{6000}.$$

It turns out that the most singular terms appearing upon substitution of (11) in (12) are the terms $O(1/\tau^6)$. Thus, we formulate a system of the form (9), by requesting that all terms from $O(1/\tau^6)$ to $O(\tau^{32}) = O(\tau^{38-6})$ be eliminated from (12).

The singular value decomposition of $C$ is performed yielding the dimension of the null space of $C$ equal to $\dim \ker(C) = 2$. This means that there are exactly two independent expressions of the form (12) (or multiples of them), which form a two-dimensional basis set of the null space of C, i.e. two independent integrals. The expressions (12) corresponding to the two orthogonal basis vectors of *ker(C)* are:

$$\Phi_1 = 0.211944988455(y^2 + p_y^2 - 4y^3) + 0.049612730813(x^2 + p_x^2) +$$
$$+ 0.216443010189(xp_x p_y - p_x^2 y - x^2 y^2 - \frac{1}{4}x^4) - \quad (13)$$
$$- 0.207446966722 x^2 y$$

$$\Phi_2 = 0.035960121414(y^2 + p_y^2 - 4y^3) + 0.342416478352(x^2 + p_x^2) -$$
$$- 0.408608475918(xp_x p_y - p_x^2 y - x^2 y^2 - \frac{1}{4}x^4) - \quad (14)$$
$$- 0.480528718746 x^2 y$$

Both $\Phi_1$ and $\Phi_2$ are first integrals of the system (12). It is easy to obtain the integrals in a recognizable form. The Hamiltonian is given by:
$$H = 2.1645645936295\Phi_1 + 1.146586289822\Phi_2 \quad (15)$$

while a second integral, orthogonal to the Hamiltonian, is given by:

$$I_2 = -2.1645645936295\Phi_2 + 1.146586289822\Phi_1 \quad (16)$$

The integral $I_2$ is a linear combination of the Hamiltonian and of the integral $I_b$ given by Bountis et al. (1982). We find
$$I_2 = 0.1651752123636227(2H - \frac{12}{7}I_b) \quad (17)$$

### 3.3. Second example. ( Bogoyavlensky - Volterra Systems)

The Bogoyavlensky-Volterra B-type systems are given in normalized coordinates $u_i$ by the following set of autonomous nonlinear ODEs:

$$\begin{aligned} \dot{u}_1 &= u_1^2 + u_1 u_2 \\ \dot{u}_i &= u_i u_{i+1} - u_i u_{i-1} \qquad i = 2,...,n-1 \\ \dot{u}_n &= -u_n u_{n-1} \end{aligned} \quad (18)$$

### 3.3.1 Restrictions to the form of the integrals by singularity analysis

The r.h.s. of Eq.(18) are homogeneous functions of second degree in the variables $u_i$. The system (18) admits a special solution of the form:

$$u_i = \frac{a_i}{\tau} \tag{19}$$

where $\tau = t - t_0$ is the time near a singularity $t_0$ in the complex t-plane. The $a_i$ are solutions of the set of algebraic equations:

$$\begin{aligned} -a_1 &= a_1^2 + a_1 a_2 \\ -a_i &= a_i a_{i+1} - a_i a_{i-1}, \ i=2,\ldots,n-1 \\ -a_n &= -a_n a_{n-1} \end{aligned} \tag{20}$$

and they are given by the recursion formulas:

$$a_{k+2} = a_k - 1, \ a_1 = (-1)^n [\frac{n+1}{2}], \ a_2 = -1 - a_1 \tag{21}$$

for $k = 1,\ldots,n$. The resonances are found by the ARS algorithm (Ablowitz et al. 1980). Taking into account the recursion relations (16), the determinant of the characteristic equation turns out to have the following tridiagonal form:

$$\det \begin{pmatrix} a_1 - r & a_1 & 0 & 0 & 0 & 0 & \ldots & 0 \\ -a_2 & -r & a_2 & 0 & 0 & 0 & \ldots & 0 \\ 0 & -a_3 & -r & a_3 & 0 & 0 & \ldots & 0 \\ \cdot & & \cdot & \cdot & \cdot & 0 & & 0 \\ \cdot & & & \cdot & \cdot & \cdot & & \cdot \\ \cdot & & & & \cdot & \cdot & \cdot & \cdot \\ 0 & & & & & -a_{n-1} & -r & a_{n-1} \\ 0 & \cdot & \cdot & \cdot & 0 & & -a_{n-1} & -r \end{pmatrix} = 0 \tag{22}$$

yielding the resonances

$$r_k = (-1)^k k, \ k = 1,\ldots,n \tag{23}$$

Since there are $[n/2]$ positive resonances, there are at most $[n/2]$ polynomial integrals of the system (18). We construct these integrals by our algorithm. Let $I_n^{(m)}$ be a homogeneous polynomial integral of degree $m$ for the system (1) with $m$ variables, and $m$ even with $0 \le m \le n$. Then, we find that $I_n^{(m)}$ is defined by the recursion formula:

$$I_n^{(m)} = I_{n-1}^{(m)} + u_n^2 \sum_{k=0}^{m/2-1} [I_{n-m-1+2k}^{(2k)} \prod_{j=n-m+1+2k}^{n} u_j] \tag{24}$$

where $I_n^{(0)} = 1$ and $I_n^{(m)} = 0$ for $m > n$.

For $m = 2$, Eq(24) yields the second degree integral

$$I_n^2 = \sum_{i=2}^{n} u_i^2 + \sum_{i=1}^{n-1} u_i u_{i+1} \tag{25}$$

Similarly, for $m > 4$ Eq.(24) yields the fourth degree integral

$$I_n^{(4)} = u_n^2 \sum_{i=2}^{n-2} (u_i^2 + 2u_{i-1}u_i) + \sum_{k=4}^{n-1} [(u_k^2 + 2u_k u_{k+1}) \sum_{i=2}^{k-2}(u_i^2 + 2u_{i-1}u_i)] + \sum_{i=2}^{n-3} u_i u_{i+1} u_{i+2} u_{i+3} \tag{26}$$

In the same way we obtain the integrals of higher degree, but the formulas involved are rather cumbersome.

## 4 CONCLUSIONS

1) A new algorithm is presented which calculates explicitly first integrals of the form (2) for systems of nonlinear differential equations of the form (1), where the relevant functions are polynomial, rational or algebraic. The algorithm has, as only input, the information provided by the singularity analysis of the evolution equations (1).

2) Local series solutions around the position of a singularity $t_0$ in the complex time domain are found by means of the ARS algorithm.

3) The theorem of Yoshida (1983b, theorem 1) is used to provide constraints to the maximum degree of the terms in an integral of the form (2) for a given set of equations. The maximum degree is given as a function of the highest positive resonance in the corresponding Painlevé - Laurent series solutions.

4) The integrals are found by combining terms in such a way that the time $\tau = t - t_0$ (where $t_0$ is the position of a movable singularity) vanishes from the expression (2), in which the dynamical variables $x_i$ are replaced by their respective Painlevé - Laurent series expressions. The problem is reduced to finding non-zero solutions of a homogeneous linear system of equations with a finite number of unknowns, namely the coefficients $a_{k1,k2,...,kn}$ in Eq.(2). The solutions, yielding the integrals, are found by means of the singular value decomposition algorithm.